# Staggered dispersions: Part I. Shocliton, quantum revival and fractalization

Jian-Zhou Zhu/朱建州

[a]Su-Cheng Centre for Fundamental and Interdisciplinary Sciences Gaochun Nanjing 211316 China

**Abstract**

Alternating the signs of the frequencies for Fourier components with even and odd (normalized) wavenumbers in the nonlinear-wave, such as the Korteweg-de Vries, equations maintains a constantly drifting dispersive shock with similar solitonic oscillations on both sides, like some plasma and quantum shocks. Such *staggered dispersions* support bi-directional waves and can symmetrize and deregularize (to some degree) the nonlinear dynamics, with physical consequences that can also be reflected in the fractalization and quantum revival (Talbot effect).

## 1. Introduction

The Korteweg-de Vries (KdV) equation as a "universal" model for the dynamics of many dispersive media, ranging from water (from which the KdV paper originated [1]) to plasma waves (to which the discovery of the "solitons" were associated [2]) with continued derivations and applications of the model in specific problems [3], among others, admits marked oscillations on only one side of a shock, as evidenced by various results in the literature. This seems reasonable, for dispersive waves originating from the shock are travelling unidirectionally according to the nature of the KdV dispersion.

Shocks of real physical systems however do often present genuine "two-sided" oscillations. For example, we can find such oscillations in the ion-acoustic shock waves observed in experiments [4] and in the one-dimensional particle-in-cell simulation results of Ref. [5] (with numerical noise at even smaller scales, typical of such Lagrangian method, though). With a quantum nonconvex dispersion arising from the development of spin-orbit coupled Bose-Einstein condensates (BECs) [6, 7], they also appear in the quasi-one-dimensional dynamics [8], especially in their two-component Gross-Pitaevskii equation (GPE) simulations. These oscillations on the two sides appear to be not of that drastically different features in terms of wavelength and/or amplitude (see the nice accounts in Refs. [9, 10] for those produced by the Kawahara dispersions); they, in general, are neither from "external" forcing.

So, we propose *staggered dispersions*, which is accomplished in a cyclic box by alternating/staggering the signs of the neighbouring frequencies in the linear dispersion term of, for instance, the original KdV equation, on staggered sublattices of respectively, say, even and odd (normalized) wavenumbers in the Fourier space. [Readers with relevant background may be drawn to some remote and formal analogy with the antiferromagnetic arrangement of spins known as the *Néel state* whose staggered magnetization however is in the configuration space [11]. We observe with interest that our initial terminology, (even-odd) alternative dispersion, parallels the recent adoption of "altermagnetism" in the spintronics and magnetic materials community, which describes new variants of traditional staggered magnetization. This connection suggests "alterdispersion" as a natural extension for variants of our staggered dispersion concept. Physically, such dispersions can be viewed as an effective models combining normal dispersions of different orders, structurally analogous to but more complex than Kawahara's third- and fifth-order dispersions, incorporating additional higher-order terms.] Such (almost symmetrically) staggered KdV (sKdV) system supports bi-directional wave propagation, thus new forms of balance leading to novel patterns including the emergence of stable solitonic (anti-)shock-like structures, from the initial data originally used by Zabusky and Kruskal [2] ('ZK' hereafter), to be presented in this work. Neighbouring even and odd wavenumbers are not really "twins", so that the minor asymmetry

*Email address:* jz@sccfis.org (Jian-Zhou Zhu/朱建州 )

between them can lead to effects that need to be corrected, which will be completed with the introduction of the "boy-girl-twin" dispersions of the same amplitude but opposite signs.

On the other hand, the notion of soliton [2, 12, 13] and multi-soliton solutions, or the analogue in a periodic problem, acquires a kind of mathematical preciseness in the inverse scattering method in terms of the spectrum of the corresponding Lax-pair operator (see also, e.g., Refs. [14, 15] and references therein for, respectively, infinite-gap theory and the physical modeling of soliton gas with a special ansatz of the gap distribution in the large-gap-number limit). But, the term "soliton" is frequently used in a more general sense, as done in many studies of nonintegrable and even dissipative systems, and many other terms ending with "-on", following the particle physics convention, have been introduced to denote various solitonic structures.

Also fundamental is the fractalization and quantum revival (FaQR), associated to the Talbot effect, linear and nonlinear ([16, 17, 18, 19, 20, 21, 22, 23]. So, well-controlled models are needed to help shedding additional lights on such or more general structure-formation issues. Our interest on FaQR is partly due to the obvious shock-antishock structures in the quantum-revival solutions with piece-wise constant initial data whose graphs, with the Gibbs phenomena from finite Fourier-mode representation, look roughly similar (but essentially different) to the above-mentioned (anti)shock structures carrying minor oscillations.

Since our sKdV model supports bi-directional waves, more introductions on relevant properties follow. Linear waves of the same wavelength travelling oppositely can form standing-wave mode which might enhance the nonlinear interaction. This could be the physical reason causing the singularity of the Boussinesq equation, even in the so-called "good" case:

$$u_{tt} = u_{xx} + (u^2)_{xx} - u_{xxxx} \tag{1.1}$$

whose Hamiltonian formulation, integrability and blow-up have been established [24, 25] (see Ref. [26] for a possible explanation between the formal paradox between the inverse-scattering linear problem and the original nonlinear blow-up).

In Farmakis et al. [23], cubic nonlinearity is used to replace the quadratic nonlinearity in Eq. (1.1), showing non-Gibbs oscillations and even somewhat strong spikes. We will also see strong spikes in some of the sKdV results. No evidence of blow-up have been clearly obtained, which remains for further studies. Here, our focus is on the discovery and its possible reason of the emergence of (pseudo-)periodic patterns with shocks of two-sided similar oscillations from such presumably nonintegrable systems.

Below, we will introduce in Sec. 2 the staggered dispersions, and, the variational and Hamiltonian formulation, with particularly the discussions of on-torus invariants and quasi- and almost-periodic tori, and, FaQR. Sec. 3 contains the basic numerical analyses and tentative physical applications in modeling the ion-acoustic shocks, including both the cases with sinusoidal and step-function initial data and, correspondingly, the respective (dispersive) shock formation and FaQR. Sec. 4 concludes the work with expectations.

## 2. Theoretical formulation

The KdV equation for the physical variable ('velocity' hereafter) $u$ parameterized by a dispersion coefficient $\mu$ writes

$$\partial_t u + u\partial_x u + \mu\partial_x^3 u = 0 \tag{2.1}$$

which, in a periodic interval of length $L_p$ normalized to be $2\pi$ for convenience of theoretical formulation, reads in the Fourier $k$-space

$$(\partial_t - \mu\hat{i}k^3)\hat{u}_k + \hat{i}\sum_{p+q=k} q\hat{u}_p\hat{u}_q = 0 : \tag{2.2}$$

with the Fourier coefficient

$$\hat{u}_k(t) = \int_0^{2\pi} u(x,t)\exp\{-\hat{i}kx\}dx/(2\pi) =: \mathscr{F}\{u(x)\}(k,t) \tag{2.3}$$

with $\hat{i}^2 = -1$ and the complex conjugacy $\hat{u}_k^* = \hat{u}_{-k}$ for real $u$,

$$u(x,t) = \sum_k \hat{u}_k \exp\{\hat{i}kx\} =: \mathscr{F}^{-1}\{\hat{u}\}(x,t) \tag{2.4}$$

with appropriate properties depending on our requirements on the behaviors of the series (in general, we assume convergence, thus simply the equality with the Fourier expansion, but for special cases involving discontinuities, such as the FaQR below, slightly more careful usage of the notation will be applied for caveat).

We will be studying the sKdV equation, with $\mathrm{mod}\,(k,2) = [(-1)^k+1]/2$ effecting a pseudo-differential operator to re-group even and odd wavenumbers to be assigned opposite-sign dispersions:

$$\partial_t u + u\partial_x u + \mu\partial_x^3\Big[\mathscr{F}^{-1}\big\{\,\mathrm{mod}\,(k+1,2)\mathscr{F}\{u\}(k)\big\} - \mathscr{F}^{-1}\big\{\,\mathrm{mod}\,(k,2)\mathscr{F}\{u\}(k)\big\}\Big] = 0. \tag{2.5}$$

*2.1. The definitions and equations*

The modification of the dispersion term in Eq. (2.5) of the KdV dispersion $\hat{D}_k = -\hat{i}\mu k^3\hat{u}_k$ is, put in words, simply separating the even and odd-$k$ modes and reverse the sign of the dispersion of one of the branches (in the fashion of configuration-space staggered spin in the Néel antiferromagnet) with appropriate assumptions of the Fourier series; that is,

$$\hat{\tilde{D}}_k = (-1)^{\,\mathrm{mod}\,(k,2)}\hat{i}\mu k^3\hat{u}_k \tag{2.6}$$

which is really what will be used below. But, for slight generalization and for a glance at its formulation in physical space, some more tedious but trivial elaborations (mostly just for introducing some more notations for later usage) of the implicit pseudo-differential operators involved follow.

In $x$-configuration space, the more general, in the sense of unequal ${}^e\mu$ and ${}^o\mu$ below, even-odd staggering of the dispersion term $D$ is modified to be $\tilde{D} = {}^e\tilde{D} + {}^o\tilde{D}$, where

$${}^e\tilde{D}(x) = -\hat{i}\,{}^e\mu\sum_k(2k)^3\hat{u}_{2k}\exp\{\hat{i}2kx\} = -\hat{i}\,{}^e\mu\sum_k k^3\,{}^e\hat{u}_k\exp\{\hat{i}kx\} = {}^e\mu\partial_x^3\,{}^eu, \tag{2.7}$$

with ${}^e\hat{u}_k := \mathrm{mod}(k+1,2)\hat{u}_k$; and, similarly for the odd component,

$${}^o\tilde{D} = -\hat{i}\,{}^o\mu\sum_k(2k+1)^3\hat{u}_{2k+1}e^{\hat{i}(2k+1)x} = -\hat{i}\,{}^o\mu\sum_k k^3\,{}^o\hat{u}_k\exp\{\hat{i}kx\} = {}^o\mu\partial_x^3\,{}^ou, \tag{2.8}$$

with ${}^o\hat{u}_k := \mathrm{mod}(k,2)\hat{u}_k$. That is, we have

$$\partial_t u + u\partial_x u + ({}^e\mu\partial_x^3\,{}^eu + {}^o\mu\partial_x^3\,{}^ou) = 0 \tag{2.9}$$

with

$${}^eu = \mathscr{F}^{-1}\big\{\,\mathrm{mod}\,(k+1,2)\mathscr{F}\{u\}\big\} \text{ and } {}^ou = \mathscr{F}^{-1}\big\{\,\mathrm{mod}\,(k,2)\mathscr{F}\{u\}\big\}, \tag{2.10}$$

as used in Eq. (2.5); or, in $k$-space, the dispersion function $\omega(k)$ in Eq. (2.6), $\hat{\tilde{D}}_k = \Omega(k)\hat{u}_k$, being slightly generalized,

$$\partial_t\hat{u}_k - \hat{i}\Omega(k)\hat{u}_k + \hat{i}\sum_{p+q=k} q\hat{u}_p\hat{u}_q = 0 \tag{2.11}$$

with

$$\Omega(k) = k^3[\,{}^e\mu\,\mathrm{mod}\,(k+1,2) + {}^o\mu\,\mathrm{mod}\,(k,2)]. \tag{2.12}$$

${}^e\mu$ and ${}^o\mu$ may be chosen to be independent, but we will start with $\mu = {}^e\mu = -{}^o\mu$ (thus the conventional KdV dispersion $D = {}^e\tilde{D} - {}^o\tilde{D}$) which is almost symmetrical and which is used in the numerical analysis in Sec. 3 where purely symmetrical staggered dispersion with further modification to correct the minor asymmetry made in Sec. 2.3.2 will also be used to show “improvements”.

*2.2. The variational principle and Hamiltonian formulation*

Gardner's [27] KdV formulation in $k$-space can be carried over for sKdV, but for convenience of further theoretical discussions on travelling waves and invariant tori, we formally lay out the similar results in $x$-space.

Obviously, for any variable $v$, the even-projection operators $\mathscr{E} :\ v \to {}^e v$ and, similarly, $\mathscr{O}$ are *linear*, satisfying

$$\mathscr{E}\mathscr{E} = \mathscr{E},\ \mathscr{O}\mathscr{O} = \mathscr{O} \text{ and } \mathscr{E}\mathscr{O} = \mathscr{O}\mathscr{E} = \mathscr{N} :\ v \to 0. \tag{2.13}$$

*Commutativity* of $\mathscr{E}$ and $\mathscr{O}$ with differentiation operators is obvious. And just as the derivatives of $v$, ${}^{e/o}v$ should be treated as independent variables derived from $v$ in functional calculations. The variation of ${}^{e/o}v$ comes from that of $v$. Introducing $\phi$ with $\phi_x = u$ (thus ${}^{e/o}\phi_x = {}^{e/o}u$ — subscripts of $\phi$ denoting partial differentiations), we have the least-action variational principle

$$\delta \int L dx dt = 0 \tag{2.14}$$

with the Lagrangian density

$$L = \phi_x \phi_t/2 + \phi_x^3/6 - {}^e\mu({}^e\phi_{xx})^2/2 - {}^o\mu({}^o\phi_{xx})^2/2. \tag{2.15}$$

Eq. (2.9) follows the Euler equation

$$\frac{\partial}{\partial t}\frac{\partial L}{\partial \phi_t} + \frac{\partial}{\partial x}\frac{\partial L}{\partial \phi_x} - \frac{\partial^2}{\partial x^2}\Big(\frac{\partial L}{\partial\, {}^e\phi_{xx}} + \frac{\partial L}{\partial\, {}^o\phi_{xx}}\Big) = 0. \tag{2.16}$$

[The above equation can be derived by introducing a $\phi$ variation (which causes the corresponding variations of ${}^e\phi_{xx}$ and all that functions in $L$) and perform the standard direct computations, with the application of the properties (especially the *linearity* and *commutativity* with respect to differentiations) of the operators $\mathscr{E}$ and $\mathscr{O}$, and, $\mathscr{F}$ and $\mathscr{F}^{-1}$ in Eq. (2.10).] Actually, we may define the Hamiltonian

$$\mathcal{H} = \int_0^{2\pi} [\, {}^o\mu(\partial_x\, {}^o u)^2/2 + {}^e\mu(\partial_x\, {}^e u)^2/2 - u^3/6)]dx, \tag{2.17}$$

with the even and odd velocities given in the above through the Fourier expansion of $u$ after Eqs. (2.7) and (2.8) respectively. We can directly verify, with Eq. (2.11), that

$$\frac{d\hat{u}_k}{dt} = \frac{\hat{i}}{2\pi} k \frac{\partial \mathcal{H}}{\partial \hat{u}_k^*}. \tag{2.18}$$

We can also verify

$$\partial_t u = \{u, \mathcal{H}\} = \frac{\partial}{\partial x}\frac{\delta \mathcal{H}}{\delta u} \tag{2.19}$$

with Hamiltonian operator $\partial_x$ and the *Poisson bracket*

$$\{\mathcal{F}, \mathcal{G}\} = \int_0^{2\pi} \frac{\delta \mathcal{F}}{\delta u}\frac{\partial}{\partial x}\frac{\delta \mathcal{G}}{\delta u} dx. \tag{2.20}$$

In other words, with only the change of the linear dispersion term, much of the classical KdV carries over, *mutatis mutandis*.

We note that the decomposition and assignment of dispersion, and, the above variational principle and Hamiltonian formulation hold actually for quite "arbitrary" grouping and linear dispersion-assignment (except for some possibly exotic ones which might lead to mathematical difficulties) of the Fourier modes and can be carried over to continuous-$k$ case (for unbounded domain) by working with the intervals, say, $(I - 0.5, I + 0.5]$ for integer $I$s. As indicated, the models such constructed is not expected to simulate some type of physical problems universally for all kinds of initial and boundary conditions, but rather to help exposing and understanding some dynamical mechanisms in specific situations.

Further possibility of the structures in this line with the spirit of Lie's theory and others deserves examination and is left for future studies. Here, in the following two sub-sections, we merely discuss in such a context the travelling waves and quasi-periodic solutions, akin to the classical theories of the periodic KdV problem.

### 2.2.1. *Travelling waves*

The periodic KdV cnoidal wave, as the name suggests, already takes the nontrivial mathematical form (Jacobi elliptic function). With the somewhat complex pseudo-differential operator involved, the corresponding sKdV travelling wave, if exists, probably cannot be explicitly expressed by a known function. The assumed travelling wave of velocity $\lambda_1$ corresponds to the special solution to Eq. (2.11), with

$$\partial_t \hat{u}_k = \hat{i}k^3[\,{}^{e}\mu \text{ mod } (k+1,2) + {}^{o}\mu \text{ mod } (k,2)]\hat{u}_k - \hat{i}\sum_{p+q=k} q\hat{u}_p\hat{u}_q = -\hat{i}\lambda_1 k\hat{u}_k. \tag{2.21}$$

Actually, it is direct to check that still two more of the KdV invariants,

$$\mathcal{E} = \int_0^{2\pi} u^2 dx \text{ and } \mathcal{M} = \int_0^{2\pi} u dx, \tag{2.22}$$

are conserved accordingly by sKdV. That is,

$$\{\mathcal{E}, \mathcal{H}\} = 0 = \{\mathcal{M}, \mathcal{H}\}. \tag{2.23}$$

So, we see that the above wave can be realized by the critical set of the KdV flow specified by

$$\frac{\delta\mathcal{H}}{\delta u} - \lambda_1 \frac{\delta\mathcal{E}}{\delta u} = 0. \tag{2.24}$$

Whether including $\mathcal{M}$, vanishing or not, does not matter here, as in KdV [28, 29].

Note that, introducing the Galerkin regularization/truncation (Gr), i.e., keeping only modes of a finite set of wavenumbers, makes it possible to write down some explicit solutions to Eq. (2.21), just as in other Gr-models [30], which however is not relevant to the purpose here: for extended information, we will further introduce some notions similarly developed also for the Gr-systems [30, 31], but, even for interested readers, it is not necessary to run into the details of the latter parallel studies until the concluding discussion in Sec. 4 for an in-depth and critical comparison with the counter-balancing effect of nonlinear Gr-dispersion.

### 2.2.2. *On-torus invariants*

No other sKdV invariants beyond $\mathcal{H}$, $\mathcal{E}$ and $\mathcal{M}$ can be found so far (again, the unconventional pseudo-differential operator appears to hinder the classical approaches, say, along the Lie theory [32], to symmetry and invariance). However, as we will see below, similar multi-soliton analogue like the KdV case is indicated by the sKdV numerical results with periodic boundary condition, which somehow calls for a setup resembling the infinitely many invariants of the integrable KdV.

Actually, for any integral functionals $\mathcal{I}_\tau$ ($\mathcal{I}_1 = \mathcal{E}$) defining the tori by the critical sets of the sKdV flow with

$$\frac{\delta\mathcal{H}}{\delta u} - \sum_\tau \lambda_\tau \frac{\delta\mathcal{I}_\tau}{\delta u} = 0, \tag{2.25}$$

it is straightforward to show the on-torus invariance, i.e., $d\sum_{\tau\neq 1}\lambda_\tau\mathcal{I}_\tau/dt = 0$, or $\{\sum_{\tau\neq 1}\lambda_\tau\mathcal{I}_\tau, \mathcal{H}\} = 0$. Chosen to be mutually Poisson commuting (if possible), each $\mathcal{I}_\tau$ then should Poisson commute with $\mathcal{H}$, so that

$$d\mathcal{I}_\tau/dt = \{\mathcal{I}_\tau, \mathcal{H}\} = 0, \tag{2.26}$$

i.e., on-torus invariance individually. [The KdV integral invariants in Ref. [27] mutually Poisson commute, but here $\mathcal{I}_\tau$ can be quite arbitrary, thus not commuting unless chosen to be so.] We may let $\mathcal{I}_\tau$ be quadratic in $u$. This way, it could be possible to have solutions to Eq. (2.21) with the right hand side replaced by more general $\hat{i}\omega(k)\hat{u}_k$, thus the possibility of quasi- and almost-periodic tori, with $\omega(k)$ solving the *generalized eigenvalue problem*[1] corresponding to Eq. (2.18). Then we would have a setup resembling that for the KdV finite-gap theory, though we do not know the corresponding operator, not even for its existence, as in that KdV Lax pair.

[1] When $\omega(k)$ is a $k$-independent constant, we have the classical eigenvalue problem, but $k$-dependent $\omega(k)$, i.e., the (infinite-order) matrix with different eigenvalues, corresponds to a generalized case.

Nevertheless, we tend to believe that the on-torus-invariance scenario is relevant to the numerical results to be presented, which might be generic for (nonintegrable)[2] conservative systems associated to the concept of pseudo-integrability (introduced also in the paralleled work for the Gr-models [30, 31]). The latter notion is associated with the patterns, to be discovered numerically, with solitonic structures accompanied by (much) weaker disordered components, indicating the corresponding whiskered tori. For those Gr-models [30, 31], such patterns are called "longulence", with the corresponding solitonic structures called "longons". Now the similar scenario related to the corresponding whiskered torus is that the stable manifold is responsible for the solitonic structures with apparent periodic character (probably quasi- or, in the case of $K \to \infty$ and infinite frequencies, almost-periodic), and that the unstable manifold for the chaotic components: the orbit can somehow escape with perturbations from the unstable "whiskers" and come back to the stable manifold. We thus have to introduce a new term, "pseudo-periodicity", for the combination of periodicity of the solitonic structures and the chaoticity of the disordered components.

Each such torus carries a set of invariants specific to it. For each realization, finding out the "right" on-torus invariants can be challenging, but not necessarily completely untraceable. For example, each Hamiltonian flow $u_t = \partial_x \frac{\delta \mathcal{I}_\tau}{\delta u}$ should reflect some aspect(s) of the structure of the tori or the physical feature (such as some kind of symmetry). Indeed, in a broader or more fundamental sense, together with Refs. [30, 31], this is also an effort to understand the mathematical and physical essence of "integrability", or more practically definite, to find the possibility of bridging the conventional notions of "integrability" and "nonintegrability".

### 2.3. *On the dispersion, shock and oscillations*

We will see that the dispersive quantization and fractalization, leading to pulses from, say, piecewise constant data corresponding to the physical (anti)shocks, are consistent with the oscillation scenario.

#### 2.3.1. *Linear dispersion and dispersive quantization/fractalization*

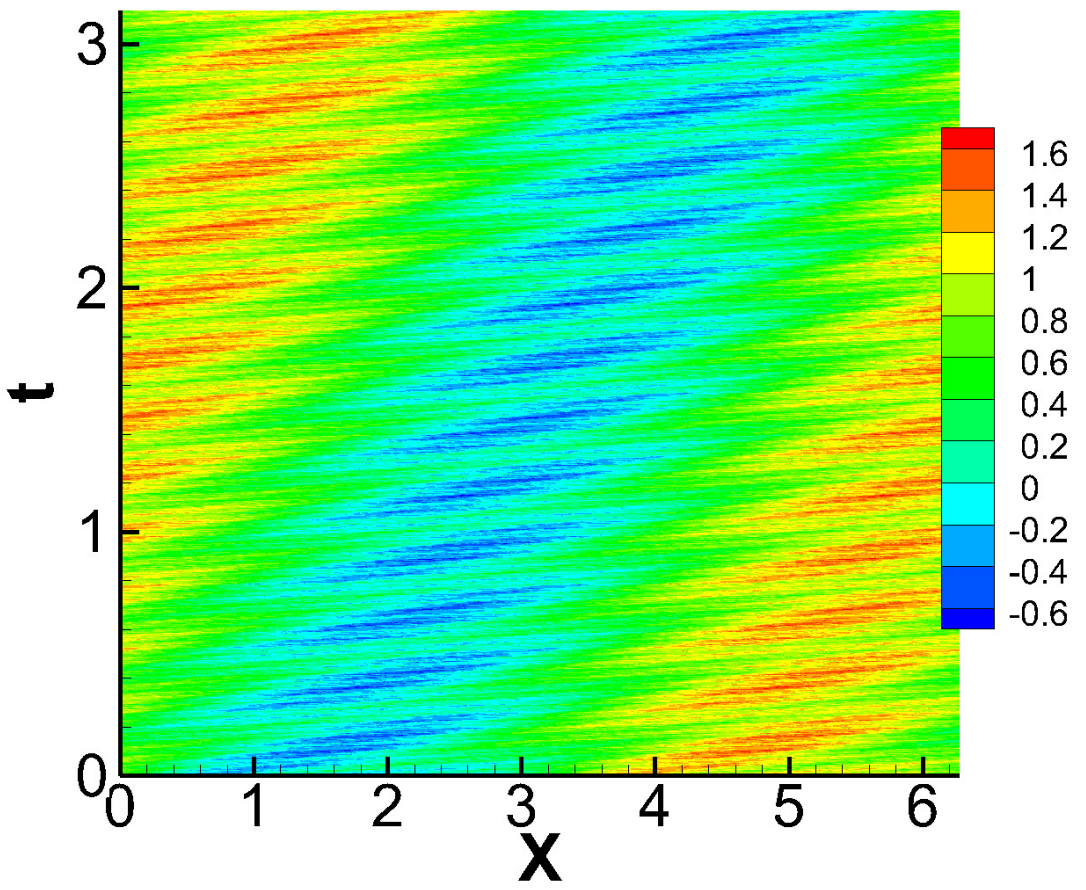

Figure 1: LsKdV space-time contours/carpets from the simple step-function initial data (2.28), with ${}^o\mu = 1 = -{}^e\mu$.

Without the nonlinear advection/dispersion term, we can write down the solution,

$$u(x,t) \sim \sum_k \hat{u}_k(t) e^{\hat{i}kx} \text{ with } \hat{u}_k(t) = \hat{u}_k(0) e^{\hat{i}\Omega(k)t}, \tag{2.27}$$

[2]We use the term "(non)integrable" in the usual sense associated with the inverse scattering method, but, actually, the meaning of integrability in general is not really clear and is a subject of research.

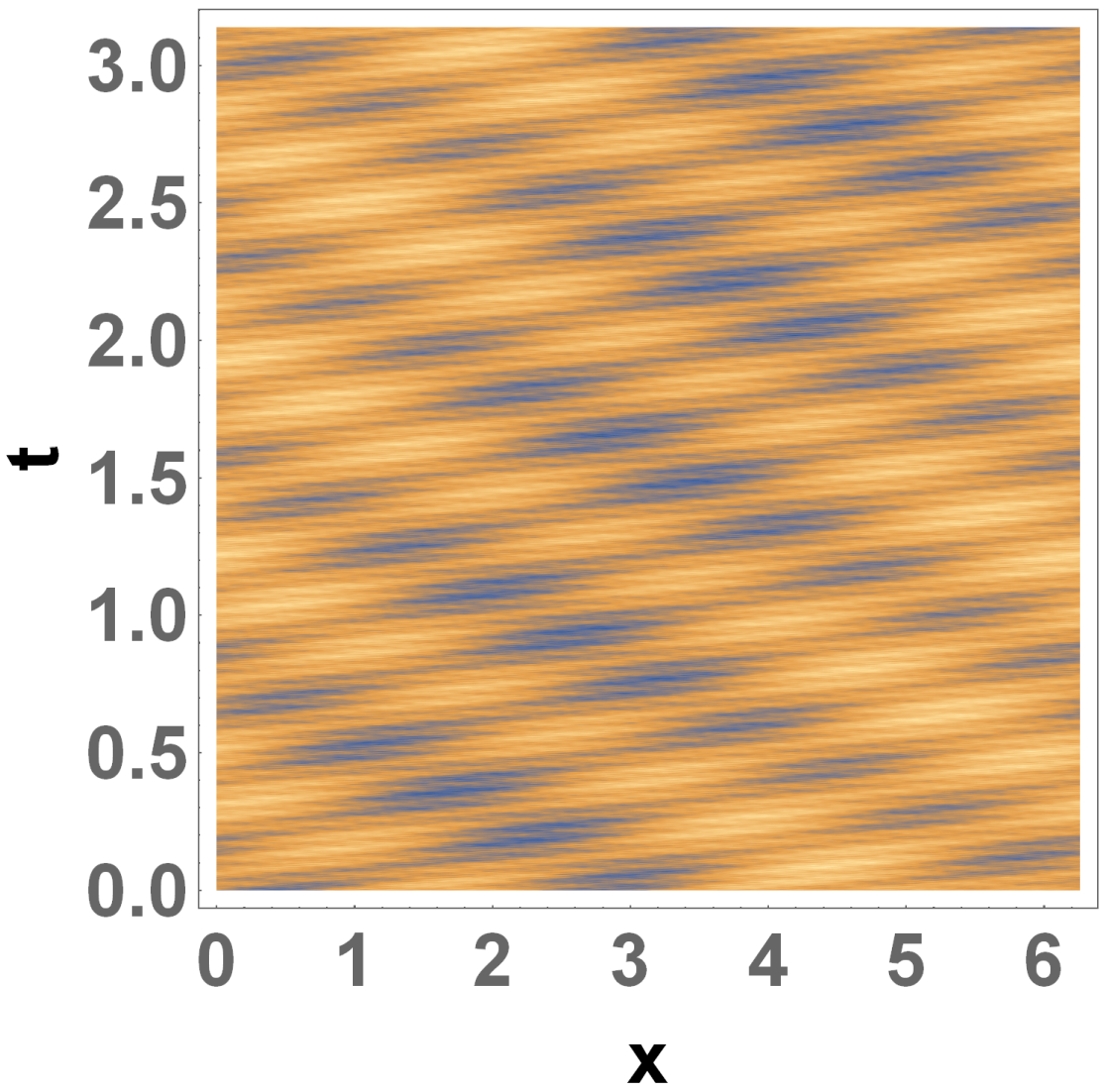

Figure 2: LsKdV space-time contours/carpets from $u(x,t) = 0$ for $x/(2\pi) \in (0, 1/8) \cup (3/8, 1/2)$, $= 1$ for $x \in (1/8, 3/8) \cup (1/2, 1)$, with ${}^o\mu = 1 = -{}^e\mu$.

to the Cauchy problem of the linear dispersion, here the linearized sKdV (LsKdV) modifying the linearized KdV (LKdV), but the result is highly nontrivial, presenting FaQR which have rich physical and mathematical background, tracing back to the Talbot effect, both linear and nonlinear, and extending to many models [16, 18, 20, 19, 22, 21, 23]. From the various familiar dispersion functions, of polynomials or not [22, 23], besides our $\Omega$, the problem is clearly of number-theoretic feature for the characters of $u(x)$ with $t$ being rational or irrational (relative to $\pi$). Our angle of view is slightly of dynamical nature. We are more concerned by how the structures or patterns are driven or formed, although further manipulations of the dispersion function appear useful for number-theoretic studies along this line in the future.

We start with the unit-step-function initial data

$$u(x,0) = \begin{cases} 0, & 0 < x < \pi, \\ 1, & \pi < x < 2\pi. \end{cases} \text{ or, } \hat{u}_k(0) = \begin{cases} \frac{\hat{i}}{\pi k}, & k \text{ odd}, \\ \frac{1}{2}, & k = 0, \\ 0, & k \neq 0 \text{ even}. \end{cases} \tag{2.28}$$

The values at the discontinuity locations, also for the case in Fig. 2, are not essential, but are taken to be the average, $1/2$, consistent with the Fourier analysis [20]. The corresponding results of Refs. [20, 21]), with ${}^o\mu = 1$, carry over with no necessity of any change, because modes of even $k \neq 0$ are vanishing, resulting in no differences between KdV and sKdV: Fig. 1 is already given by Ref. [20], but we will offer more remarks a bit later.

Actually, the following theorem for LKdV has been established [20, 21]:

**Theorem 1.** *Let* $\frac{p}{q} \in \mathbb{Q}$ *be a rational number with* $p$ *and* $q$ *having no common factors. Then the solution to the initial-boundary value problem at time* $t = \pi\frac{p}{q}$ *is constant on every subinterval of the form* $\frac{\pi j}{q} < x < \frac{\pi(j+1)}{q}$ *for* $j = 0, \ldots, 2q-1$.

**Remark 1.** *The proof in Ref. [20] is based on the expression of the Fourier expansion of* $u(x,t)$ *itself. We feel that it can be more intuitive to prove the above theorem, and similarly also relevant results in Ref. [21] on the characterization of the region of constancy, by computing* $\partial_x u(x,t)$ *and analyzing the vanishing or nonconvergent conditions, which can also be carried over, mutatis mutandis, to establish the corresponding analytical results for our LsKdV here (with the Weyl and Kummer sums etc. changed accordingly). However, since there is no essential differences in the mathematical nature, we defer such otherwise boring discussions to another communication concerning the effects on particle transports which are sensitive to such details as shocks, oscillations and fractals. Actually, as will* *be remarked again, our nonlinear sKdV results are also in a sense close to those of KdV, with differences though. So, focusing on the shock-antishock structure and the associated oscillations, we do not bother going into such detailed quantitative differences here. Presenting the scenario is sufficient for our purpose here.*

Related to our theme of dispersive (anti)shocks and oscillations, the major wave with $|k| = 1$ travels with unit velocity, as can be seen directly from the overall tilted pattern in the carpet of Fig. 1. Olver [20] noticed such "discernible wave that moves across the interval with unit speed". Actually, all details can be explained by the wave velocity $\Omega(k)/k = k^2$: the wave of $|k| = 3$ is responsible for the intermediate tilted stripes of red and blue (separated by green) colors whose slope/velocity (reciprocal) is seen to be 1/9, and fine bands inside each stripe corresponding to the wave of $|k| = 4$ of slope $1/16$ are also visible with our bare eyes. The pattern looks, very roughly, like the dispersive (anti)shock, and we will see in Sec. 3.4 that such linear dynamical features persist in the nonlinear regime with well preserved shock-antishock structure from the initial data.

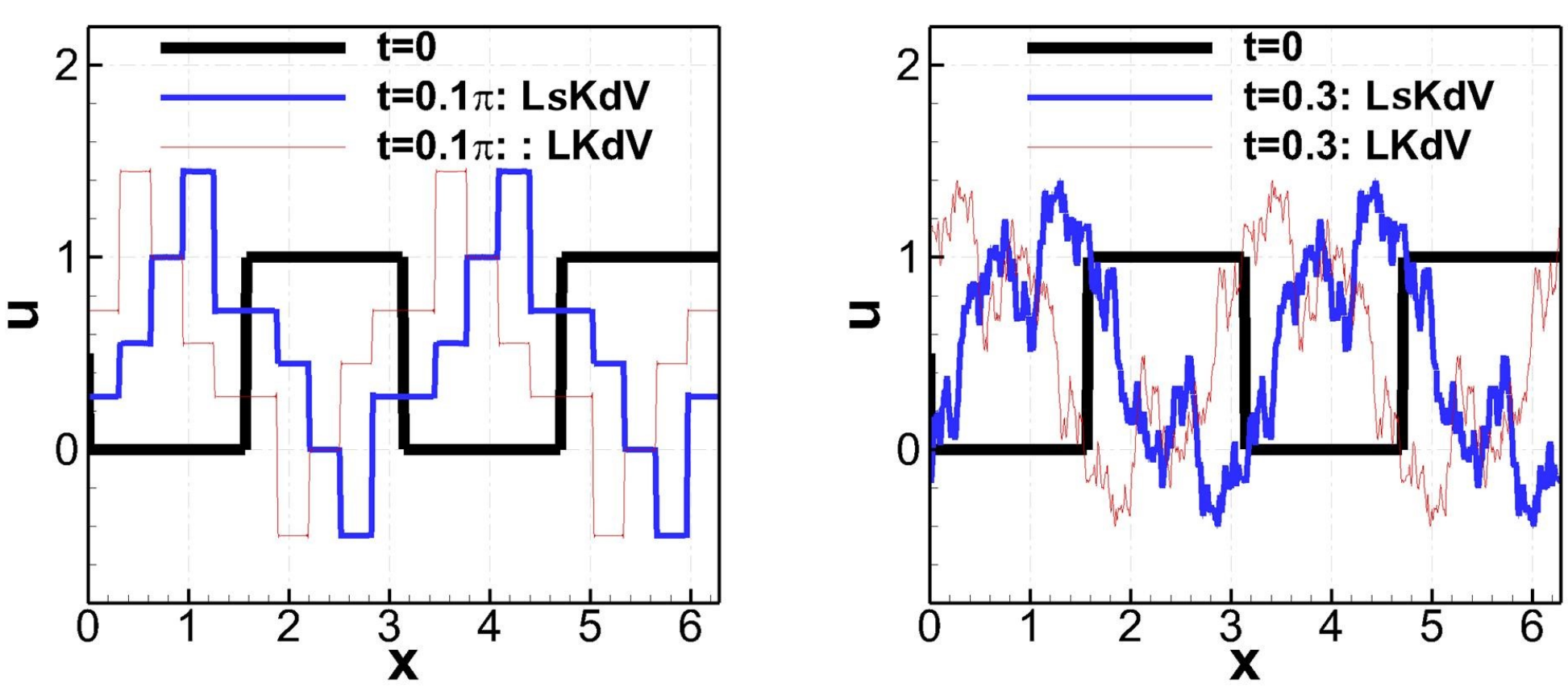

Figure 3: Quantum revival (left) and fractalization (right) of LsKdV and LKdV.

The FaQR features extend to other initial piecewise-constant data, as partly showcased in Fig. 3 with $\mu = {}^{o}\mu = 1 = -{}^{e}\mu$, snapshots of the piecewise-constant graph at the initial time $t = 0$ and respectively a rational and irrational times (with respect to $\pi$). Note that FaQR present infinitesimally fine structures and that the irrational number as a rational multiplicative of $\pi$ can only be approximated by a rational number with computer truncation, so we definitely should have sufficiently small grid sizes both in time and space discretization to accurately represent the results, similarly for the nonlinear ones to be presented later: the simple method we used to see whether the numerical errors or resolution limits influence the conclusions is making sure that there be no essential differences when those sizes are halved.

How LsKdV differs from LKdV of course depends on the initial data. Although, except for the locations, it is hard to detect with bare eyes any essential differences in the fine structures in Fig. 3 between LsKdV and LKdV, the associated number-theoretic aspect [21] is different and preliminary numerical tests show they in general are different in the shocks and fractals, resulting in amplified differences in the transports of particles. Also, just as in other nonlinear models, the corresponding persistence of FaQR should be expected in our sKdV. It is expected the differences between the sKdV and KdV dynamics would have interesting consequences on particle (density) transports.

In any case, we have seen that quantization and fractalization still (very) roughly keeps the shocks (or probably closer to "undular bores"), introducing "oscillations", quantized or fractal.

*2.3.2. Correcting the minor asymmetry in the neighboring frequencies*

In the above staggering scheme *almost symmetrically* supporting bi-directional waves, there is still *minor asymmetry* effect from the differences between the neighbouring even and odd wavenumbers. Such weak asymmetry increases with the decrease of the wavenumbers, which is intuitively obvious: the relative difference between wavenumber $n = 1$ and 2 should be larger than that between 10 and 11, say, just as their relative differences of the values of the wavenumbers. To have “boy-girl twins” of staggered frequencies for completely symmetrical dispersion, we can further replace the adjacent wavenumbers $k = 2m$ and $k = 2m - \mathrm{sgn}(m)$ in the dispersion (2.12) with the same $k = 2m - \mathrm{sgn}(m)/2$. That is, $\Omega(2m) \to \Omega(2m - \frac{\mathrm{sgn}(m)}{2}) \leftarrow \Omega(2m - \mathrm{sgn}(m))$, i.e.,

$$\Omega(k) = \hat{i}[k + \mathrm{sgn}(k) \bmod (k,2) - \frac{\mathrm{sgn}(k)}{2}]^3 [\, {}^{e}\mu \bmod (k+1,2) + \, {}^{o}\mu \bmod (k,2)]. \tag{2.29}$$

Such a correction of the linear term does not affect the conservation laws and variational formulation, and we will see that the shock propagation indeed can be nicely corrected to be of constant speed (Fig. 9 below).

## 3. Numerical analysis

Lacking a systematic method of analysis, we will particularly take the pains to carefully observe the structures of the sKdV results from the specific sinusoidal initial data used by ZK, for the possibility of evidence or clues to the physical understanding and a potential theory of pseudo-integrability towards bridging integrability and nonintegrability. Such initial data for the periodic problem has the nice properties that it is cleanly composed of a single mode of smallest $|k|$ and that it nicely demonstrates the well-understood dynamics dominated by the nonlinear term, i.e., the shock formation process up to $t_B$ (see below) before the dispersion enters to take effect. A remarkable result is the emergence and maintenance of shock-antishock structure carrying pulses, which is somehow reminiscent of the (artificial) Gibbs phenomena in ideal-shock or piecewise-constant graphs represented by finite Fourier modes and which is followed by the sKdV results of FaQR from piecewise-constant data.

### *3.1. Even-odd decompositions*

Note that the KdV even/odd dynamics

$${}^{e/o}\partial_t u + \, {}^{e/o}(u\partial_x u) + \, {}^{e/o}\mu \, {}^{e/o}\partial_x^3 u = 0, \tag{3.1}$$

which can be obtained by simply collecting from Eq. (2.2) the even/odd modes, are “non-symmetric” in the sense that the triadic interactions in ${}^{e/o}(\hat{i} \sum_{p+q=k} q\hat{u}_p \hat{u}_q)$ corresponding to the nonlinear term are composed of the “even-even”, “even-odd” and “odd-odd” ones which result in, non-symmetrically, “even”, “odd” and “even” $k$s, respectively. For example, we may remove all the odd-$k$ modes to have pure even-dynamics, but there is no nonlinear pure odd-dynamics.

Following ZK, we performed simulations with the same setup as theirs, that is, starting from the initial longest-wavelength profile $u_0 = \cos(\pi x)$ and performing the direct numerical simulation with periodic boundary condition over [0,2), but now with the standard pseudo-spectral method; $\mu = \delta^2$ with $\delta = 0.022$ in Eq. (2.1).

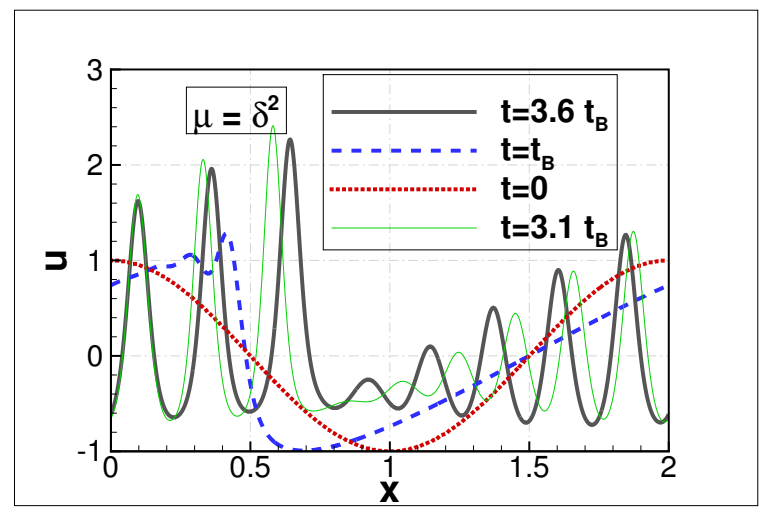

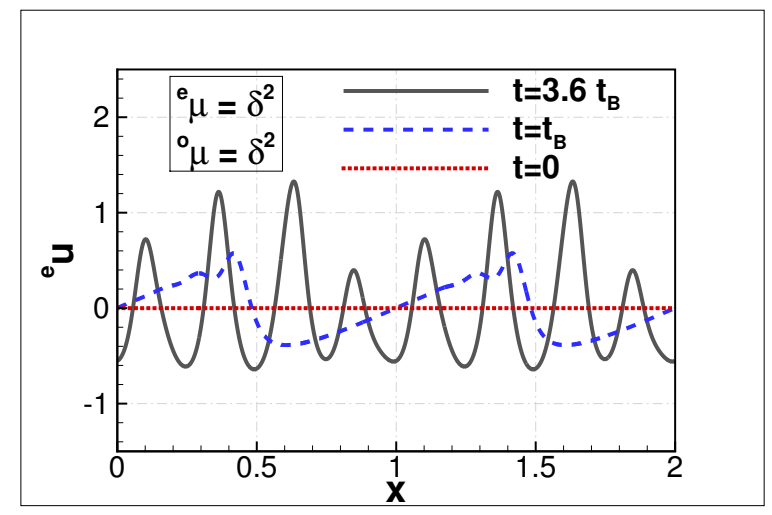

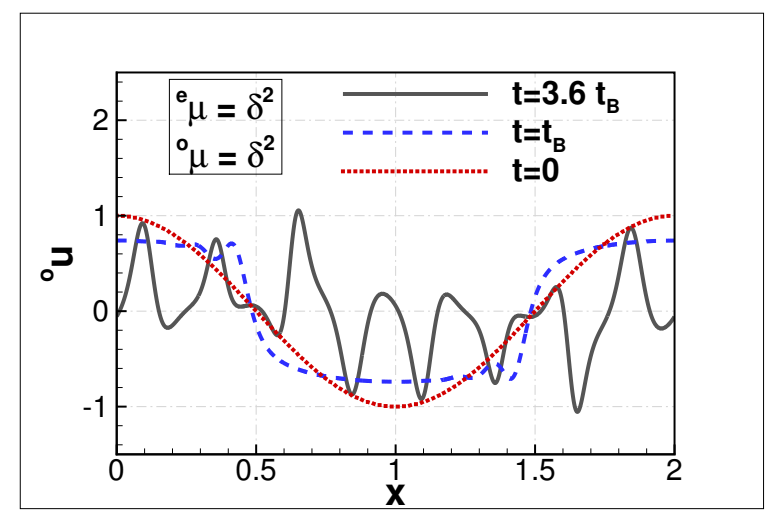

Figure 4: The KdV $u$ (after ZK) and the corresponding even and odd components. The additional line at $t = 3.1t_B$ is for showing the signature of the ninth weakest soliton (rather than the radiating signal [13])

Fig. 4 presents the velocity profiles, precisely those of ZK, and their even/odd components defined before, with the wavenumbers normalized accordingly to be integers. Some observations follow. Consistent with the initial condition, the corresponding even-velocities are actually of unit period, i.e., ${}^{e}u(x,t) = {}^{e}u(x+1,t)$, and odd-velocities of “unit anti-period”, in the sense that within the period $L_p = 2$, ${}^{o}u(x,t) = -{}^{o}u(x+1,t)$. ${}^{e}u$ tends to form shocks at both $x_B = 1$ and $x'_B = 3/2$ at the inviscid-Burgers blow-up time $t_B$ ($= 1/\pi$ now), before evolving into soliton-like structures, and ${}^o u$ to shock and anti-shock (or steep "kink", as used to describe the topological soliton of the sine-Gordon equation [33]), respectively at $x = 1$ and $x = 3/2$. Once they evolve into solitary pulses, ${}^e u$ and ${}^o u$ present similar features in their patterns, which however is not the case for the staggered-dispersion models to be presented below. Visually, from the comparison of their space-time contour plots (with still somewhat trackable space-time characteristics though) to that of the undecomposed $u$, they look less solitary than the ZK "solitons". Such characteristics are of course case specific, but still helpful. What appears remarkable is that ${}^{e/o}(u\partial_x u)$ tend to produce (anti)shocks at both $x_B = 1$ and $x'_B = 3/2$, even starting from the null field for ${}^e u$. The oscillations all emerge and develop, as already indicated by the nascent ones at $t_B$ and other subsequent ones (not shown), behind the (anti)shocks, with the linear waves of both branches supposed to propagate backward with phase velocities $-k^2$. The uni-directionally dispersive KdV waves destruct the shock into solitons, but the sKdV (almost) symmetrically staggered bi-directional dispersions of the waves can work to compensate each other's destruction, forming a different type of balance to maintain the shock (formed at $t_B$ as in the KdV case before the dispersion starts to take effect), actually to enrich the latter to be *shocliton*, which will be demonstrated below.

### 3.2. *Even-odd staggered dispersion: shocliton*

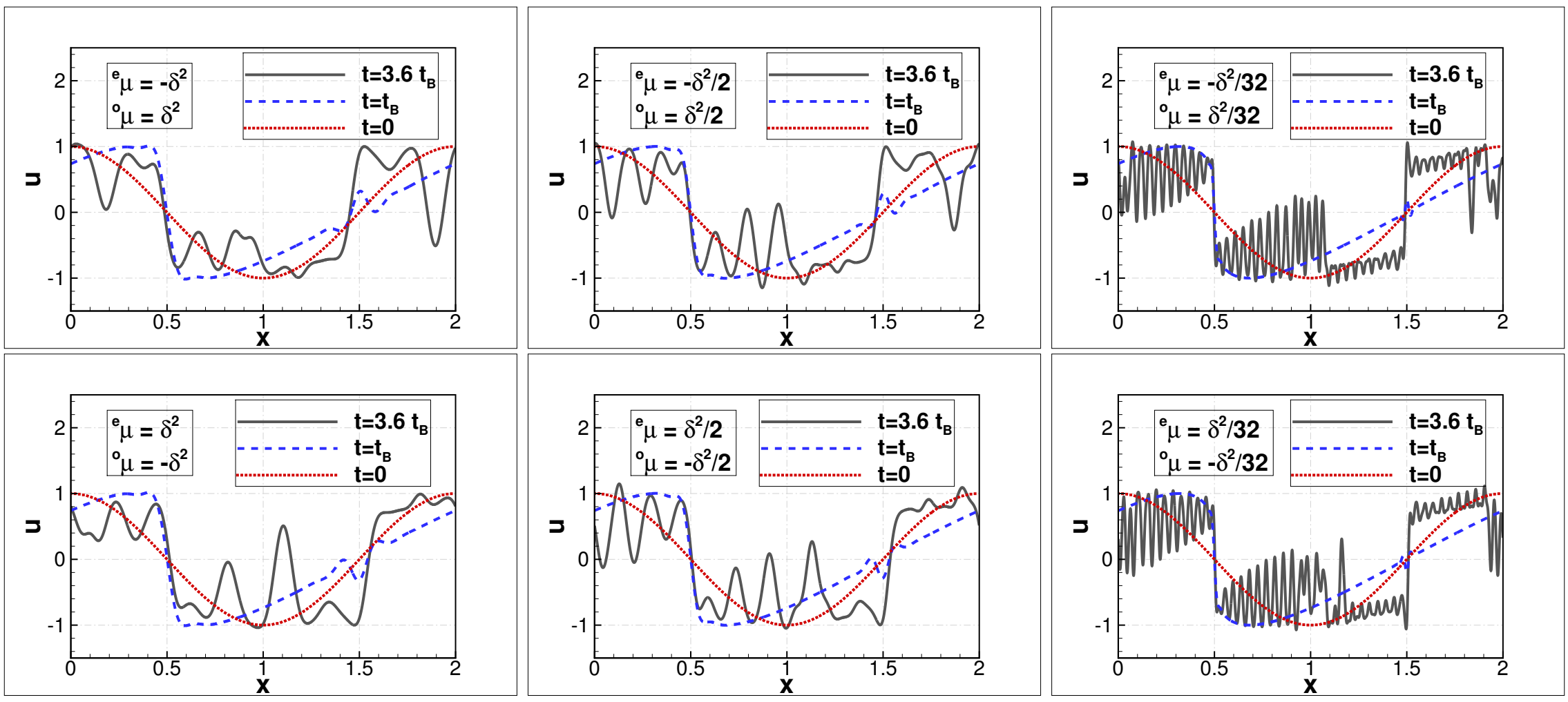

Figure 5: $u$-profiles of sKdV with different parameters.

Fig. 5 presents the velocity/$u$-profiles of the (even-odd-alternating-dispersion) sKdV equation, showing, just as the classical KdV, increasing (with decreasing $\mu = \pm\, {}^e\mu = \mp\, {}^o\mu$ from left to right) oscillations on both sides of the (anti)shocks. So, we have seen not only the objective of mimicking the two-sided oscillations observed in some plasma and quantum shocks but also the singular behavior as an indication of nonconvergence to the classical shock described by an entropy solution. We reiterate that such two-sided oscillations for given $\mu$ are not the Gibbs phenomena but "physical".

Fig. 6 presents ${}^{e/o}u$ of the case with ${}^e\mu = -{}^o\mu = \delta^2$. [The other cases are accordingly similar and not shown.] As expected, the profiles are close to the corresponding ones in Fig. 4 for KdV at $t \le t_B$, but ${}^{e/o}u$ are quite different at

$$t_{ZK} = 3.6 t_B$$

after the differences in the dispersions take more and more effect: for instance, the plateau-basin structure of ${}^o u$ of the sKdV is even strengthened (and persistent — see below), instead of broken into "simple" solitary pulses in the fashion of the KdV case.

Note that besides the shock at $x = 1/2$ as in KdV, the other anti-shock emerges at $x = 3/2$, and on both sides of the respective shock live the oscillations of close features (Fig. 5). Drastically different amplitudes of the oscillations around the respective shock can present (Fig. 5), but not always (Fig. 7 for $\mu = \delta^2/8$). The overall scenario is that the oscillations are "solitary", which will become more obvious in the observation of Fig. 8 below. And, we also see slight leftward shift or slowly travelling of the (anti)shocks, due to the small difference between ${}^{o}\tilde{D}$ and ${}^{e}\tilde{D}$ coming from that of the alternating even and odd wavenumbers: as already can be seen from the comparison between the upper and lower rows of Fig. 5 and checked by the long-time contour patterns, when the signs of the even and odd dispersion coefficients are reflected to opposite signs, the (anti)shocks of the two cases travel with opposite but same-amplitude speeds. Actually, as particularly clear at $t = 6t_B$, 9.6, $12t_B$, $18t_B$ and $19.2t_B$ for instance, the (anti)shocks seem to be "unifiable" into the oscillations. It appears then reasonable to raise the notion of "(anti)shock-soliton duality" or, probably even more precisely, "(anti)shocliton" as a mixture of (anti-)shock and soliton to indicate the continuous transition between an (anti-)shock[3] and a (normal) soliton. Such a notion will be further supported and actually verified, in Part II, to be nontrivial by its spectral structure consisting components of both the classical-shock spectrum $\propto k^{-2}$ and the solitonic-pulse spectrum $\propto k^0$, which is partly motivated by studying the particle transports there.

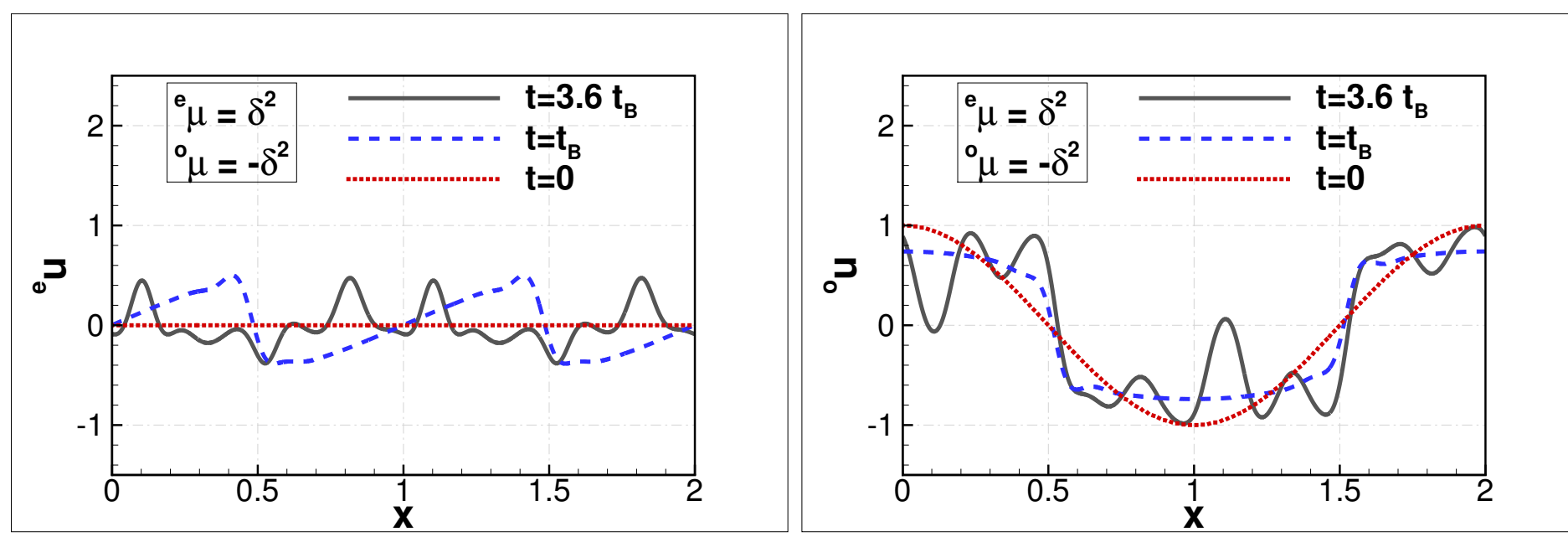

Figure 6: ${}^{e}u$ and ${}^{o}u$ of sKdV.

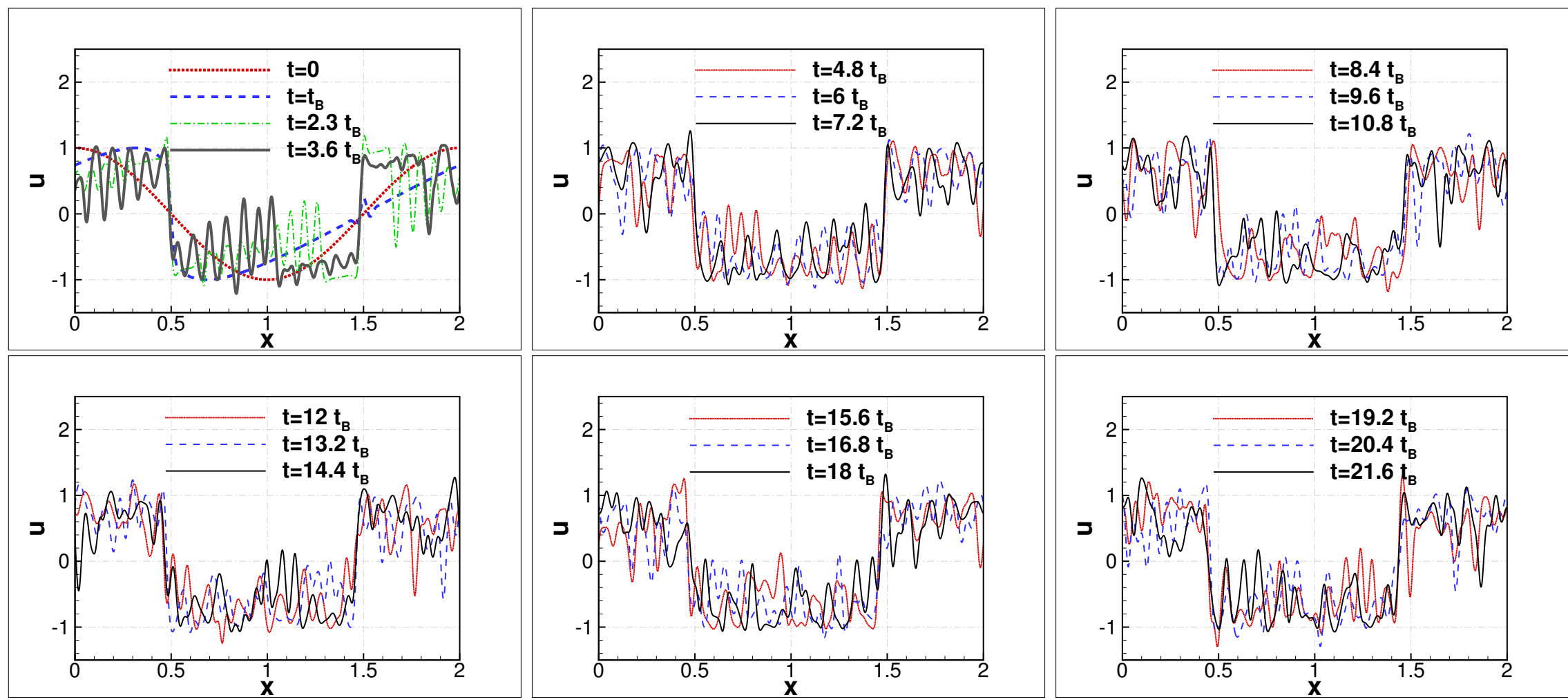

Figure 7: $u$-profiles of sKdV at various times with $\mu = {}^{o}\mu = -{}^{e}\mu = \delta^2/8$ and $\delta = 0.022$.

The obvious feature of the persistent but slowly drifting plateau-basin structure carrying the smaller oscillations on both sides of the (anti)shocks is maintained by ${}^{o}u$ (c.f. Fig. 6) in this case, while ${}^{e}u$ presents more of soliton feature, which is similarly followed by the other cases to be presented below. The conjecture, that the oscillations of such sKdV equation are solitons, appears to be supported by the contours presented in Fig. 8 where the (straight) "bars" coded by the same-level colors indicate the characteristics along which the solitary waves are travelling with collisions (interactions) resulting in some phase shifts: we indeed see there are "bars" of different slopes, indicating different velocities of the solitons, and their collisions seem to be weaker than those of the KdV equation in Fig. 4; careful inspection shows that the bars/characteristics (for solitons as conjectured by us) pass through the (anti)shoclitons which also travel at nearly constant speeds as clearly shown by their characteristics over the time up to $80t_{ZK}$,[4] supporting again the "soliton-shock duality". The pattern appears to contain some disorder, which may be described as "pseudo-periodic", with the structures at most quasi- or almost-periodic. This is similar to the "longulent states" of the Gr-systems [30, 31] mentioned in Secs. 2.2.1 and 2.2.2, with indication of the whiskered tori, and accordingly the notion of pseudo-integrability in terms of carrying a proper set of on-torus invariants [$\mathcal{H}$ and $\mathcal{I}_\tau$ in Eq. (2.25)] which precisely specify the torus.

[3]The "shock" in this note however should be distinguished from the shock "soliton" [34, 13] of viscous Burgers equation, so the notion of duality or "shocliton" is not trivial. The other solitary oscillations not of shock feature will be called "normal" solitons.

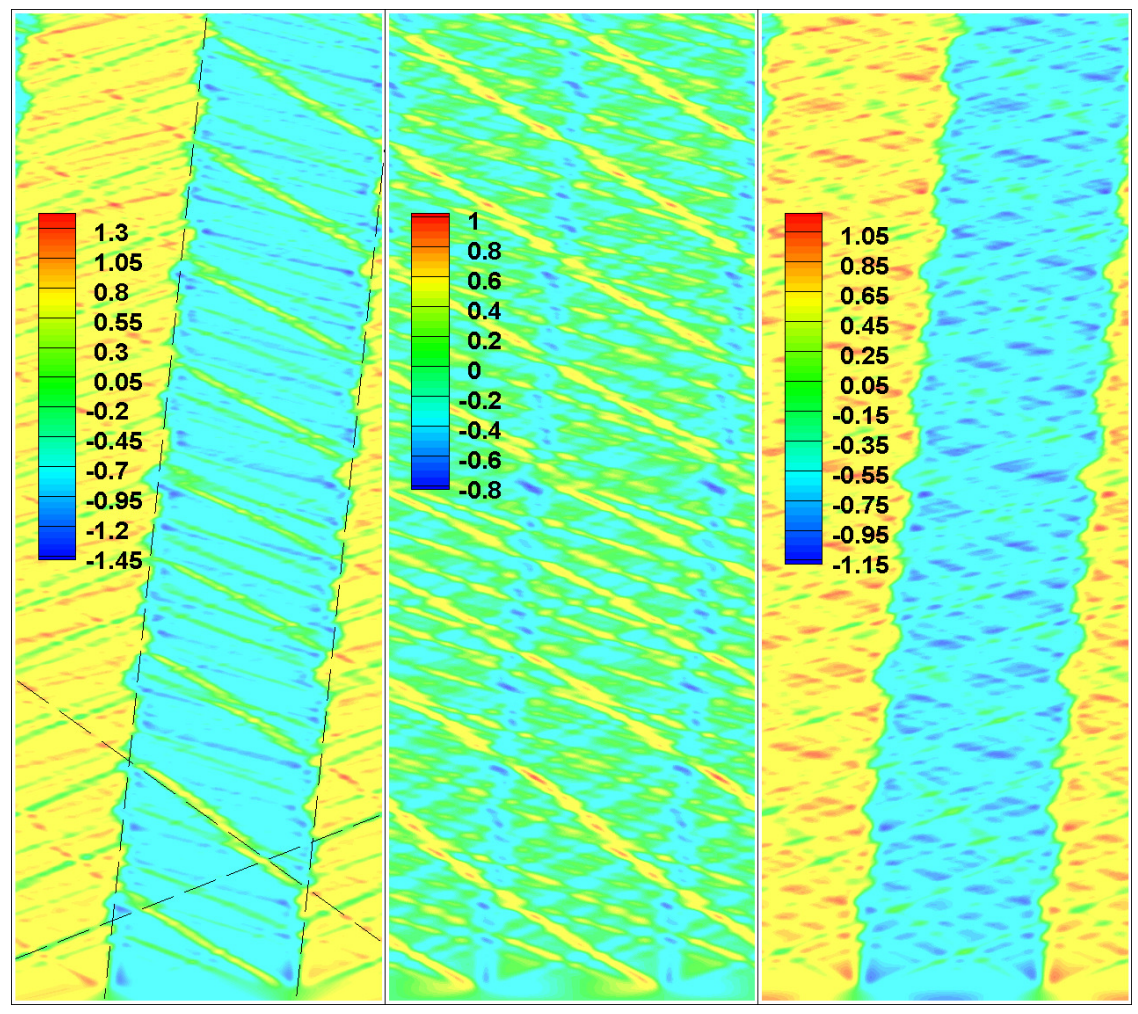

Figure 8: Contours of $u$ (left frame), ${}^{e}u$ (middle) and ${}^{o}u$ (right) for sKdV with ${}^{e}\mu = \mu = -\,{}^{o}\mu$, respectively for $t \leq 20t_{ZK}$ (with four black dashed lines added to highlight the corresponding characteristics, two longest ones of them respectively for the shocliton and antishocliton).

The shocliton-antishocliton dyad presents also in other models, including a Gr-system [30] to be further discussed in Sec. 4. The sine-Gordon equation without or with appropriate perturbations admits the "kink-antikink" pair (e.g., Refs. [37, 38]), and similarly the "kovaton" [39], which morphologically resembles our case except that, to the best of our knowledge, no oscillations around them have been studied.

From the soliton point of view, such "shoclitons" present at the top and bottom anti-directional overshoots (compared to the respective average plateau/basin profiles further away from the jumps), thus characterized to be "big but weak" ("weak" for the strength averaged over the two sides). The anti-directional overshoots on the two sides of such a shock sum up to constitute the amplitude of the (weak) soliton identity, consistent also with the specification of the classical limiting shock velocity $u_s = (u^+ - u^-)/2$ (for quadratic nonlinearity here [35]) where $u^+$ and $u^-$ are the left- and right-limit velocities of the classical ideal shock. Being "weak", the velocities of such solitons are then presumably very small, as shown by the slopes of their characteristics. Note that the (anti)shocliton phase shifts due to the collisions with solitons from both sides cannot cancel overall, due to the differences of the other counter propagating solitons.

There is slight deviation from constancy of the velocity of the "(anti)shoclitons" (not clear to our bare eyes for $t \leq 20t_{ZK}$ in Fig. 8, but marked in the left panel of Fig. 9 for $t \leq 80t_{ZK}$). We have noticed the minor even-odd asymmetry in the wave dispersion and proposed the correction recipe with Eq. (2.29) for the staggered dispersion function which, as shown in Fig. 9 (right panel) by the computation with everything else the same, works perfectly to have a corrected constant drifting velocity (which in turn indicates that the deviation from the constant drifting speed of the numerical result for the uncorrected model be genuine, not due to numerical errors): The results also show the stability and robustness of the (anti)shocliton structures and associated phenomena over long times or under perturbations, and thus their persistence in real-world scenarios, if indeed physically related.

[4]The penetration and continuation of the characteristics of the solitons across the (anti)shoclitons become much less trackable in the uniform color coding, due to the obvious jumps of the $u$-levels from the shock-property of the (anti)shoclitons.

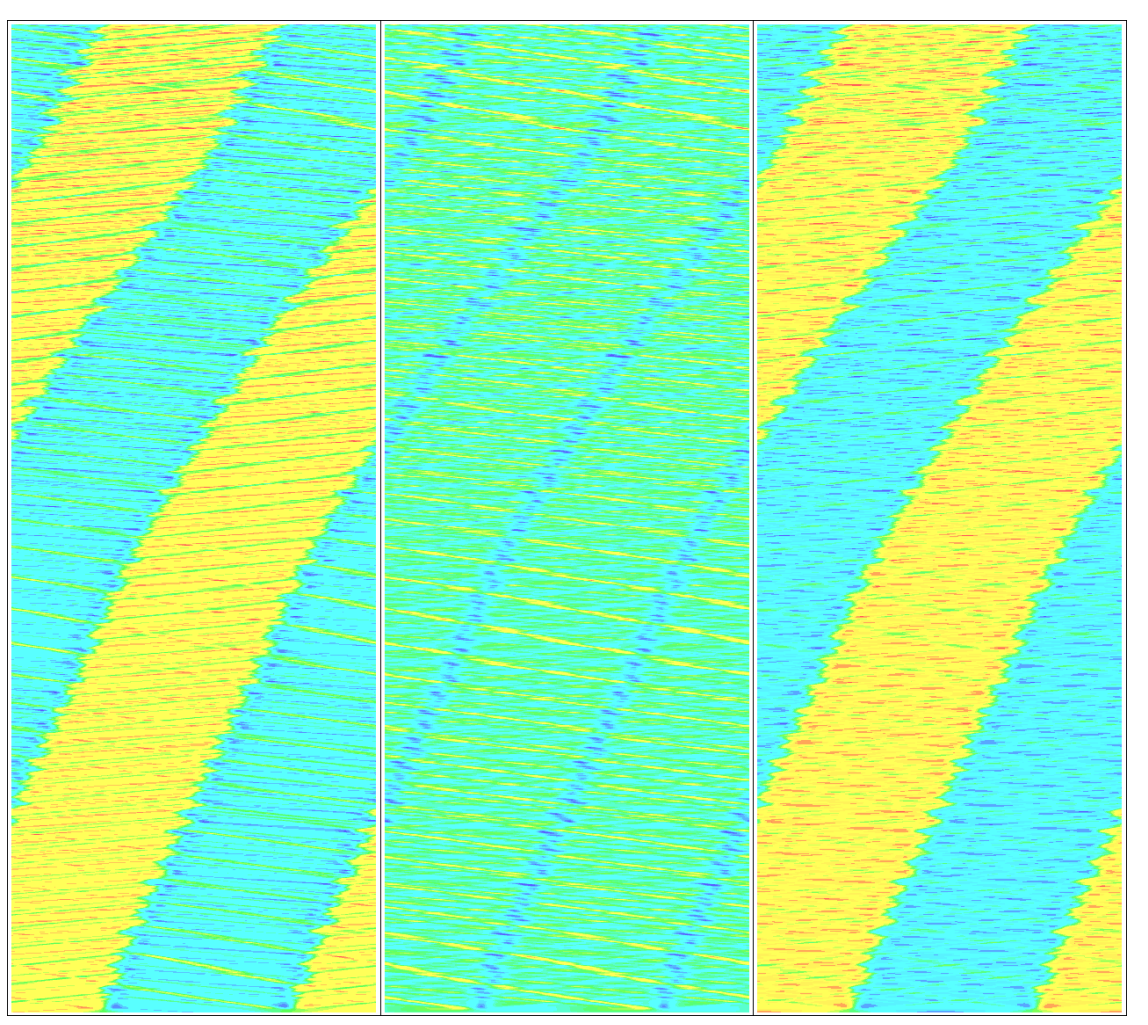
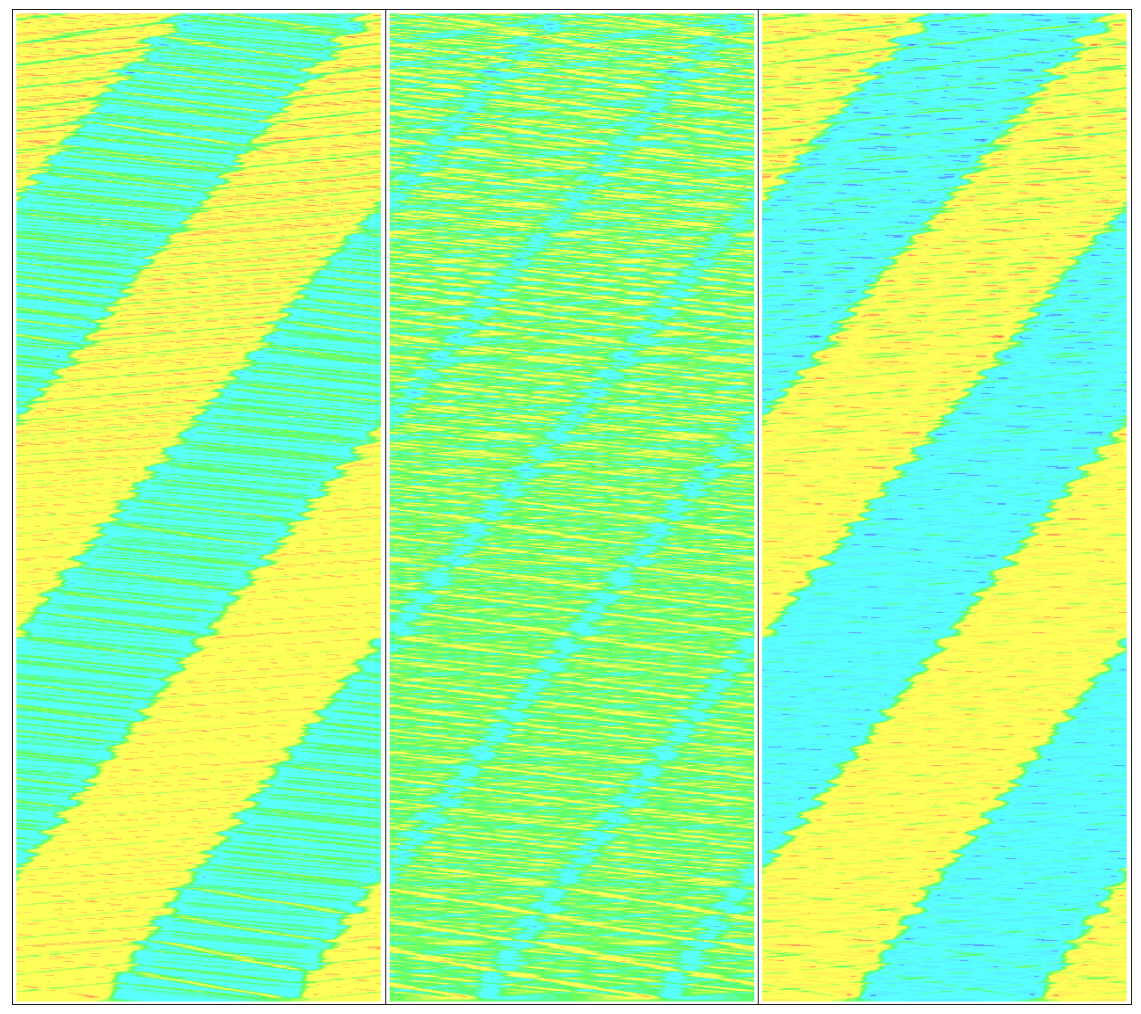

Figure 9: Contours of $u$ (left frame), ${}^{e}u$ (middle) and ${}^{o}u$ (right) for the sKdV (left panel) and accordingly those from the “corrected” sKdV (right panel), with ${}^{e}\mu = \mu = -\,{}^{o}\mu$; both for $t \leq 80t_{ZK}$.

### *3.3. Driven-damped case*

Like tuning the spin-orbit coupling of BECs [6, 7, 8] to engineer the dispersions, we can also accordingly design the latter in our sKdV models. Note that it is trivial to include the diffusive term to have our staggered-dispersion KdVB (sKdVB), and, according to the connection with quantum shocks [36], the corresponding two-sided oscillations may be associated to those found in BEC [8]. Our staggering dispersion idea for the two-sided oscillations of the shock of course belongs to the nonconvex dispersion like the Kawahara model [9, 10], but is different to the latter, most obviously in their orders. Additional specific studies are needed to see which dispersion model is more appropriate for a particular physical system.

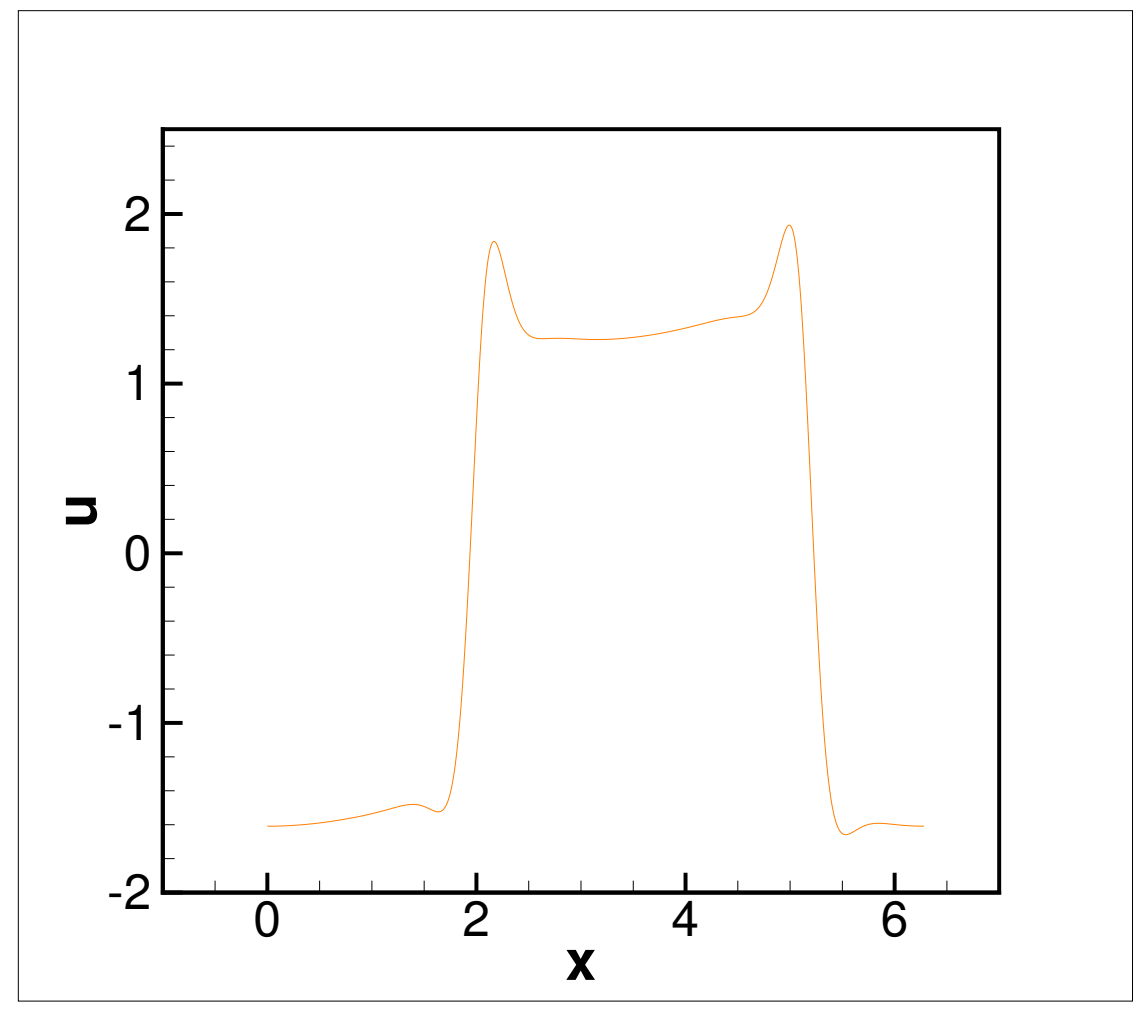

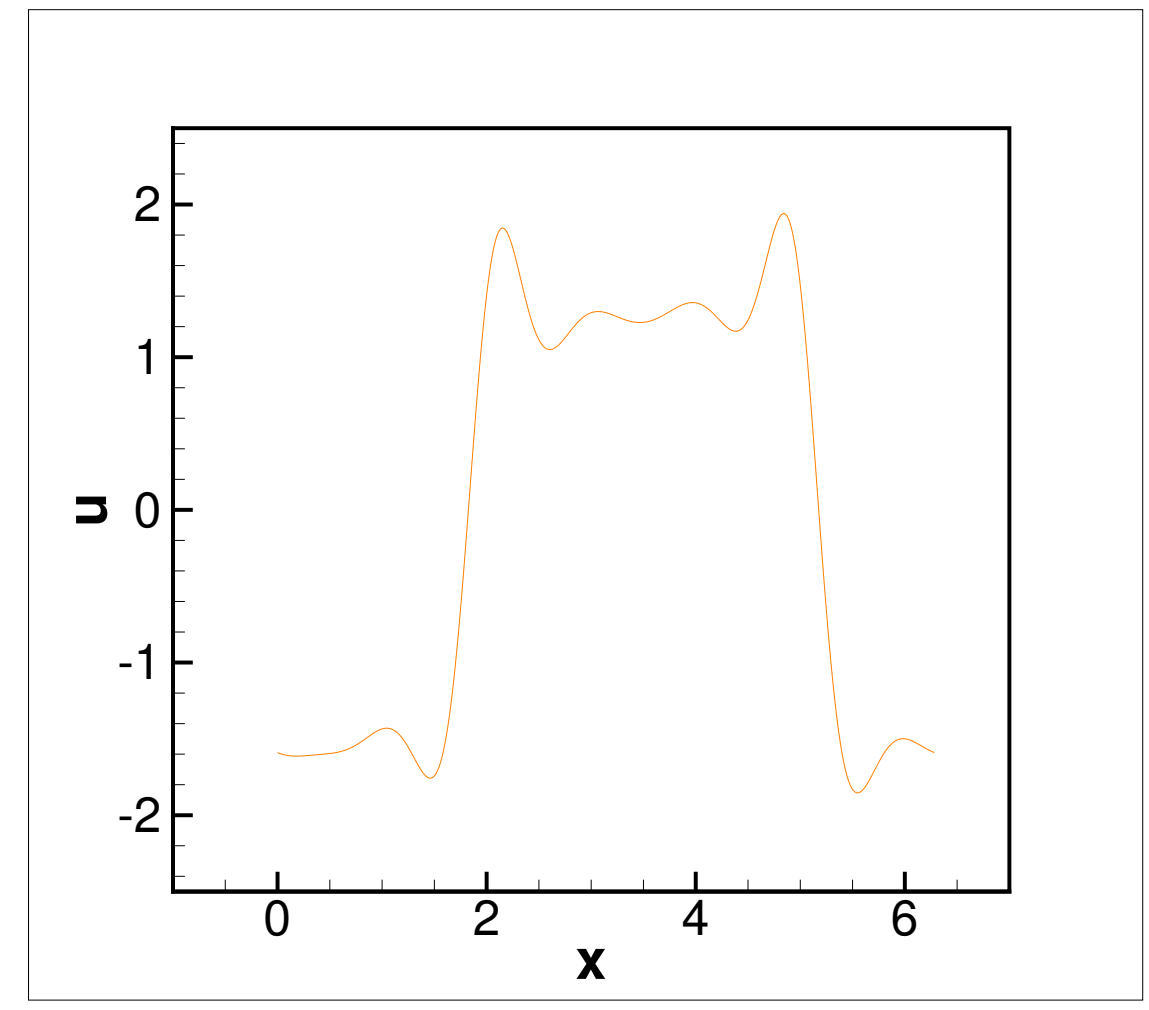

Figure 10: The sKdVB results with $m = 1$ (left: $\nu = 0.018/2$ and $\mu = 0.024/2$) and $m = 2$ (right: $\nu = 0.018/2$ and $\mu = 0.001/2$).

The sKdVB with appropriate diffusion (and forcing, if needed, say, for stationarity) can still produce the (anti)shocliton, and we should note particularly that, concerning the oscillations on both sides of the (anti-)shock, the numerical KdVB shocks with only one-sided oscillations in Ref. [4] actually could not qualitatively explain the experimental results therein, and further efforts of modelling are necessary. Now, we consider the sKdVB model

$$\{\partial_t + \nu k^2 - \mu \hat{i} k^{2m+1}[\mathrm{mod}(k,2) - \mathrm{mod}(k+1,2)]\}\hat{u}_k = \hat{f}_k - \hat{i} \sum_{p+q=k} q\hat{u}_p\hat{u}_q. \tag{3.2}$$

Using the forcing $f(x) = (\sin x)/4$, two snapshots in the (quasi-)stationary stage from simulations over a $2\pi$ period, respectively with $m = 1$ and $m = 2$ ("hyperdispersion" mentioned earlier), as presented in Fig. 10, indeed have (anti)shoclitons with two-sided oscillations, but, unlike the sKdV ones in Figs. 5 and 7, with the oscillations far away from the shock being reasonably smoothed out by the diffusion term, thus closer to the ion-acoustic shocks measured in laboratory experiment of Ref. [4]. [The even-odd asymmetry now accumulate with time until a balance is reached, leading to the systematic differences of the overshoots before and after the (anti)shock, which can be further corrected with the "boy-girl-twin" model, Eq. (2.29) applied to produce the right panel of Fig. 9: the correction can also be applied in the dissipation, but such detailed differences are not of our interest here.] Instead of fixing $m = 1$ and tuning $\mu$ and $\nu$, we have used $m = 2$ to obtain different oscillation features. Note that, like the sKdV case, the sKdVB shocliton and antishocliton in Fig. 10 do not collide but be constantly separated.

### 3.4. *Nonlinear quantum revival and fractalization*

We now test the sKdV FaQR, for good comparisons to the LsKdV FaQR in Sec. 2.3.1 (with the same setup except for the nonlinear term), to the shoclitons from the sinusoidal data in Sec. 3.2, and to the KdV results of Chen and Olver [22] (with the same setup except for the signs of the nonlinear and dispersion terms).

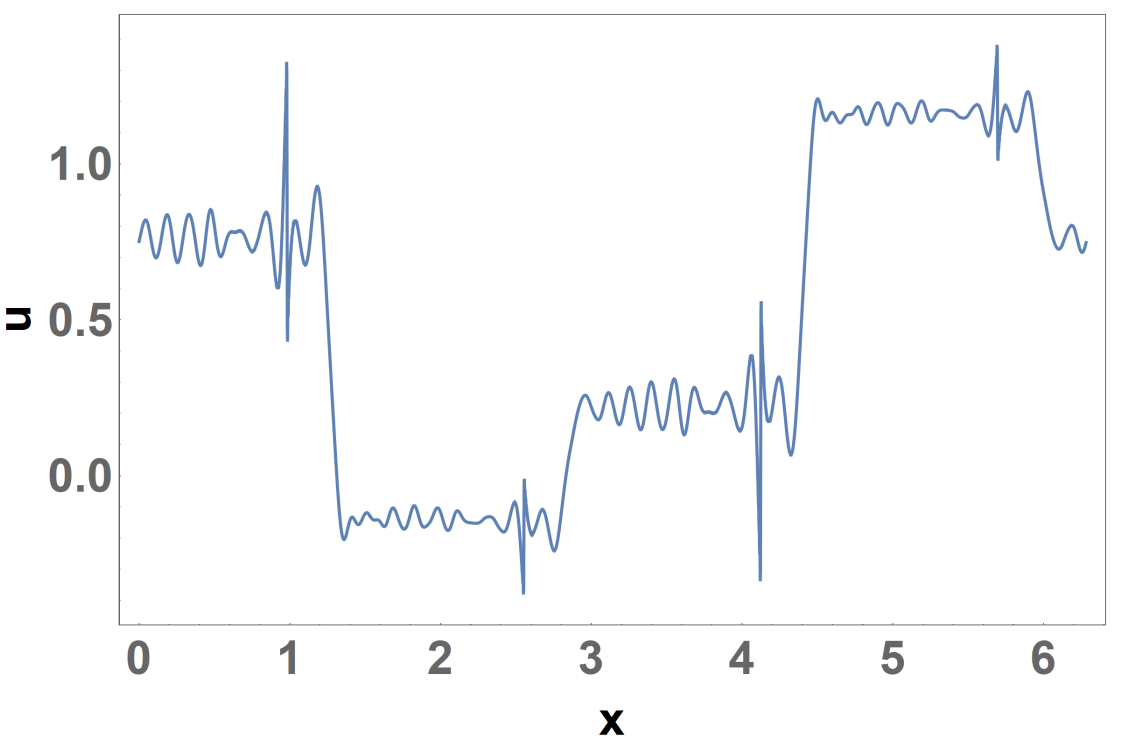

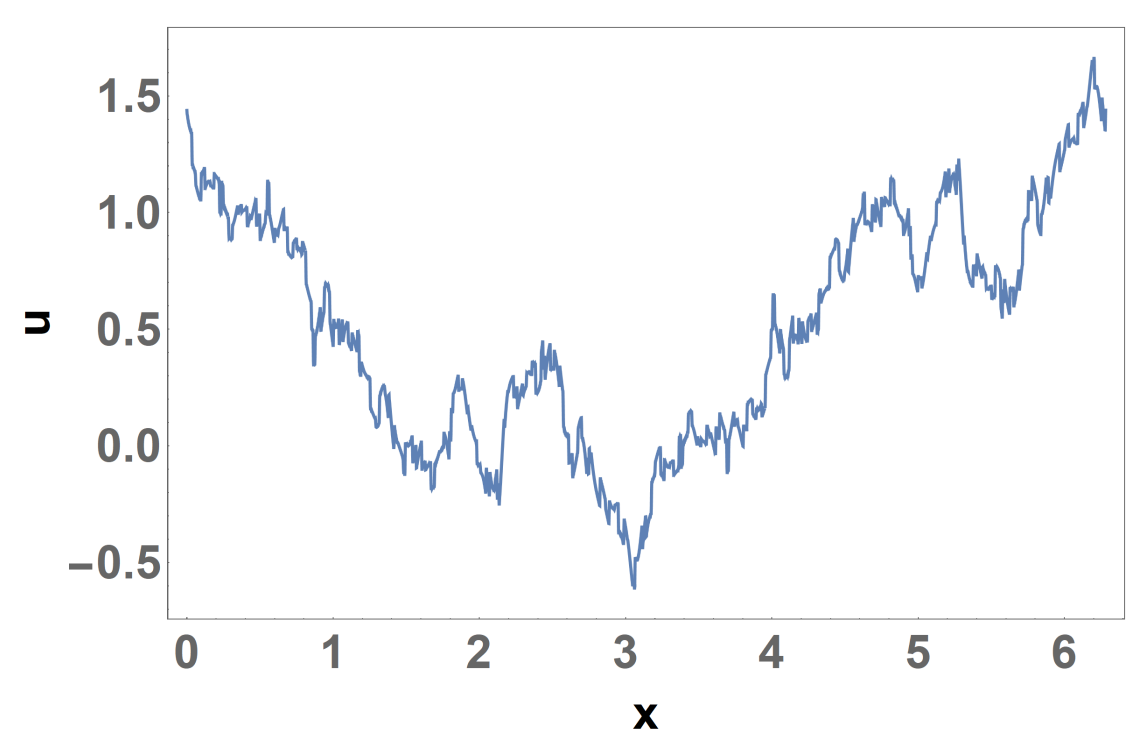

Figure 11: The sKdV quantum revival at a rational time (left) and fractalization at an irrational time (right).

Fig. 11 shows the sKdV FaQR profiles which look seemingly quite different, as in the linearized case, but overall they actually preserve the shock-antishock structure quite well, as shown by the carpet in Fig. 12. Interestingly, we have checked that such a scenario is present in the corresponding KdV results as well, with only minor differences in the details. The reason is that the shock-antishock structure, preserved also by the KdV dispersion, is mainly composed of odd-$k$ modes, with the minor oscillations about the plateau and valley/basin containing even-$k$ modes (excited by nonlinear interactions). This is different to the linear case in Sec. 2.3.1 with no nonlinear interactions to excite the even-$k$ modes, thus no difference between sKdV and KdV there. [As remarked in the introductory discussions, the quantum revival panel in Fig. 11 also presents spikes, like those in Farmakis et al. [23] who used the cubic nonlinearity to replace the quadratic one in Eq. (1.1).]

The carpet contains fine bands indicating very fast waves/solitons, i.e., the oscillations, going across the plateau and valley/basin separated by the constantly drifting shock and antishock. The shock travelling velocity is not the standard shock velocity $1/2$ as the mean of the original step function, so the (anti)shock indeed drifts additionally, actually at a speed of around $1.5$, the value as the sum of the standard shock velocity $(1/2)$ and the linear wave velocity of $|k| = 1$. The initial step data with zero mean velocity also result in additional shock drifting (not shown), similar to those starting with cosine data presented earlier. With the dispersion coefficient $1 \gg \delta^2$, the drifting here is marked compared to that in Sec. 3.2. So, aside from FaQR, the scenario is similar to what we have observed in the previous case starting from sinusoidal initial data. With FaQR, the remarkable footprints of linear waves should not be surprising.

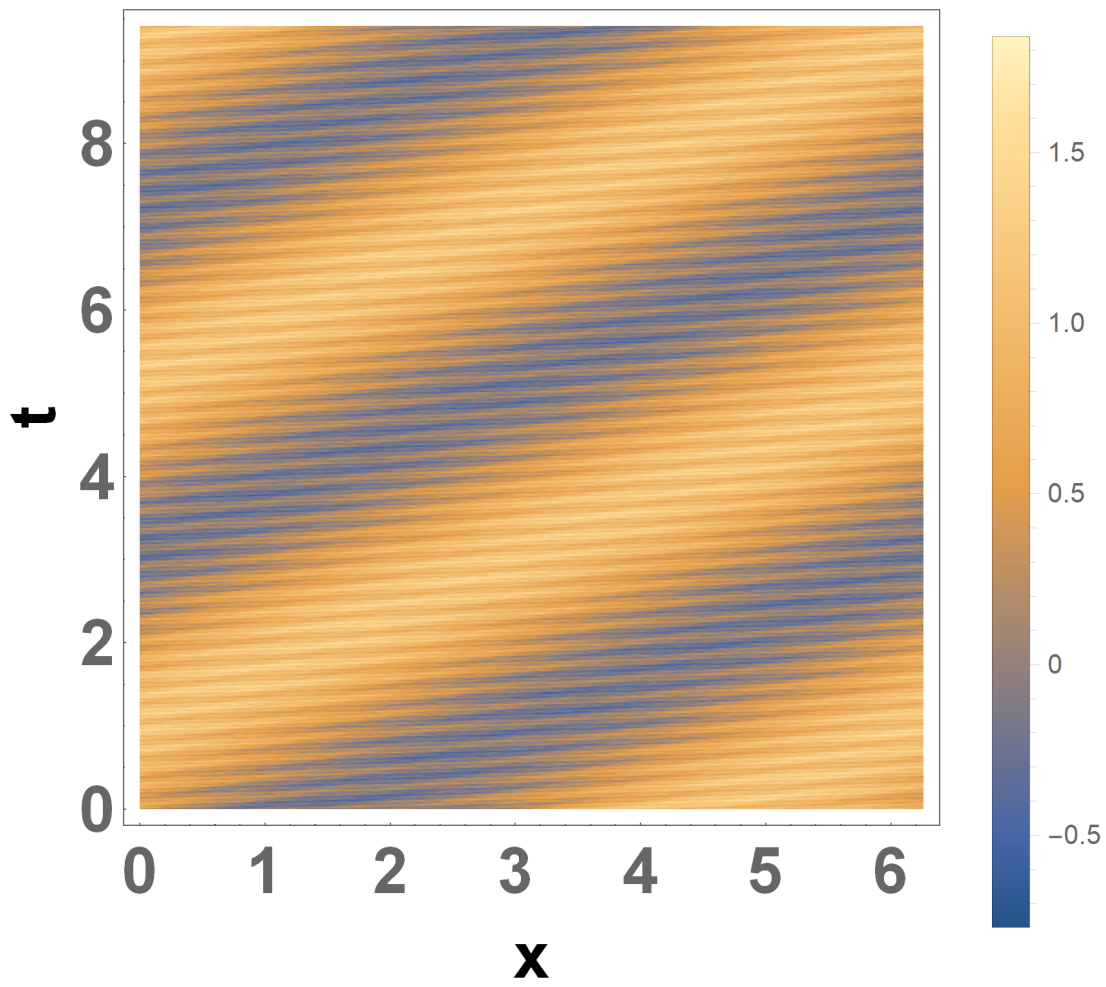

Figure 12: FaQR carpet.

So, we have found a kind of unification/consistency between sKdV soliton(-like) behavior and FaQR. Actually, both the shocliton (from the sinusoidal initial data) and the additional strong spikes (from the piecewise-constant data) indicate the staggered bi-directional dispersion not only symmetrize but also deregularize (to some degree) the nonlinear dynamics. Alternatively speaking, the uni-directional KdV dispersion has a relatively stronger regularization effect that may be reflected from its physical consequences such as the distributions of the particles transported by the KdV and sKdV flows, which is analogous to the situation that, as partly indicated in Ref. [41] in terms of the nonlinear regularization effect of reducing the compressibility related energy, more gross depletion of nonlinearity [40] presents in globally chiral (with net helicity) than in nonhelical three-dimensional (3D) isotropic turbulence. The inclusion of dissipation can be accordingly made as in Sec. 3.3 to still have the analogy, and the regularization or nonlinearity depletion analogy should apply to the incompressible turbulence: the parallel mode, like the particles transported, in the compressible case works in this respect as the detector of the regularization effect, particularly sensitive and thus convenient in the strongly compressible, especially the pressureless 3D Burgers turbulence to be used in Part II for preliminary demonstration and in a separate series of studies on helicity effects on the passive tracer and density scalar transports by (magneto-)Burgers turbulence. So, the analogy is beyond pure phenomenology and may be fundamentally essential, because it implies that, if reasonable for such perspectives, the base structures proposed in Ref. [41] of 3D turbulence present 1D soliton-like dynamics: in this sense, KdV and sKdV flows work respectively as the minimal models of helical and nonhelical turbulence.

## 4. Further discussions

This research project started from the simple observations, on the one hand, that Gardner's variational formulation and Hamiltonian structure for KdV may be equally powerful for different pseudo-differential (Fourier mode dependent) variations, such as the reassignment of linear frequencies or even the truncation of nonlinear-advection dispersion, and, on the other hand, that new dispersion by alternating the signs of the linear frequencies of the neighboring wavenumbers could well be able to simulate the wave-like pulses oscillating on both sides of the shocks, observed in some plasma and quantum mediums but not in KdV results.

Systematically carrying out the program with the sKdV model, we have not only established the above two points, with extended results such as the driven sKdVB shocks to have a broader and more concrete ground, but also discovered solitonic structures with pseudo-periodic patterns. Bearing some genericity for nonintegrable conservative systems (simultaneously supported by the discoveries of Gr-systems with the truncation resulting in nonlinear dispersion [30] — see below), the latter probably is fundamentally more important and motivating, probably needing tests on sufficiently many different initial data and in-depth analyses for further establishing the pseudo-integrability theory together with other parallel results [31]. While for interesting specific problems encountered in real physical systems, application of our model with related initial conditions, if indeed appropriate, deserves separate studies case by case.

We have also checked the applicability of the idea in the Benjamin-Ono and modified KdV (with cubic nonlinearity) models, among others, all (not shown) with similar scenarios concerning the shock-antishock and close oscillations on both sides; and, multiple nonlinearities as included in the parallel studies on Galerkin-regularized compacton and peakon models [30] in principle can also be treated with the idea of such staggered dispersions, when necessary.

The sKdV model can be viewed as adding a modification dispersion (twice the opposite of the original dispersion of the even or odd modes which is modified, otherwise zero) to the original KdV dispersion. It can be much more nontrivial to extend the staggering control idea to the nonlinear dispersions which may take exotic forms. The nonlinear advection term in the KdV equation can also be viewed as a nonlinear dispersion, and similar reassignment of the nonlinear dispersion or addition of a correction dispersion is actually realized by the familiar Galerkin truncation/regularization:

Consider for simplicity the dispersionless KdV, i.e., the Burgers-Hopf equation $\partial_t v + \partial_x v^2/2 = 0$ and the Galerkin-regularized (GrBH) dynamics [42, 30]

$$\partial_t u + u\partial_x u - {}^K g = 0 : \ {}^K g = (I - P_K)\{u\partial_x u\} = \sum_{|p|,|q|\le K<|m|}^{p+q=m} \hat{i} q \hat{u}_p \hat{u}_q e^{\hat{i} m x}. \tag{4.1}$$

In the above, use is made of the projection onto the Fourier space with $-K \le |k| \le K$ ("Galerkin space" hereafter) for some integer $K$: $P_K\{v(x,t)\} = \sum_{|k|\le K} \hat{v}_k \exp\{\hat{i}kx\}$, so $P_K$ is a time-ordered pseudo-differential operator working continuously on the dynamics, with the initial data $u_0 = u(x, 0)$ "well-prepared" in the Galerkin space. Such $u$ stays initially, and forever, in the Galerkin space which is thus dynamically "complete". Like sKdV, GrBH preserves the invariance of energy and Hamiltonian inherited from the classical KdV system [27, 43], so ${}^K g$ is a conservative, nonlinearly dispersive term ("Galerkin dispersion" or "Galerkin force"). So, we see that the reassignment of the nonlinear dispersion is realized by the Galerkin dispersion ${}^K\hat{g}_m$ which is part of the nonlinear advection/dispersion $u\partial_x u$. [We have used different notations, such as $k$ and $m$ for distinguishing different wavenumber regimes.]

The addition of opposite dispersion appears to be the common mechanism of maintaining the (anti)shoclitons. In the sKdV case, the physics mechanism should be that the unidirectional dispersion is counterbalanced by the opposite directional dispersion of waves; while in the GrBH case, the small-scale nonlinear dispersion counter-balancing component maintains also the shocliton-antishocliton in the case with quasi-piecewise-constant initial data [30]; and, of course, we can follow again Gardner's Hamiltonian formulation in $k$-space to further modify the above nonlinear dispersion ${}^K\hat{g}_m$ and to still similarly support (anti)shoclitons. Such a comparison is mutually illuminating. Indeed we had been partly led by the current work to revisit Gr-systems with new discoveries which then in turn motivated further sKdV progresses here, including to similarly propose the on-torus sKdV invariants which may lead to relevant quasi- or even almost-periodic orbits in a way similar to the integrable KdV case; and, probably most importantly, the unification of longulent states associated to presumably whiskered tori (possible to be proved by an *a-posteriori* KAM scheme in a unified way) in other presumably nonintegrable equations, such as the Gr- and hyperdispersive-systems [30, 31], has motivated on-torus or torus-specific invariants and a potential theory of pseudo-integrability towards bridging the traditional notions of integrability and nonintegrability. Note that such problems of nonintegrable systems, with the systematic solutions sounding far beyond the state-of-the-art of nonlinear mathematics though, is fundamentally interesting and practically useful (realities are usually not integrable but still with solitonic structures).

The robustness of FaQR in the sKdV model, together with the unified/consistent ubiquitous features of oscillations caused around a shock, motivate thoughts about the fundamental mechanism and applications. For example, as also hoped in Ref. [21], understanding the fractalization is an attractive, but challenging project in the field, which may benefit from the quantitative differences of KdV and sKdV; and, the particle transport, especially the aggregation/condensation, is sensitive to fractalization and shocks/quantization, which is a subject of further studies.

Combining the results of the emergence and maintenance of the shocliton-antishocliton structure (including that of sKdVB) and the FaQR, a question follows: Rather than from the initially-set appropriate singular data like the piecewise-constant ones, is it possible that FaQR emerge after some critical time when the solution becomes singular/collapsing in sKdV or other models (designed similarly or not)? The problem is reminiscent of the singularity and fractal problem of (fluid) turbulence, and we tend to believe the answer is “no” for sKdV(B), but in general it appears not impossible.